\documentclass[letter]{aa}
\usepackage{float}
\usepackage{graphicx}
\usepackage{natbib}
\usepackage{txfonts}
\begin{document}

   \title{A first glimpse at the line-of-sight structure of the Milky Way's nuclear stellar disc}

  \author{F. Nogueras-Lara
          \inst{1}             
          }

   \institute{
    Max-Planck Institute for Astronomy, K\"onigstuhl 17, 69117 Heidelberg, Germany
              \email{nogueras@mpia.de}                                 
       }
   \date{}

 
  \abstract
   {The nuclear stellar disc (NSD) is a dense stellar structure at the centre of our Galaxy. Given its proximity, it constitutes a unique laboratory to understand other galactic nuclei. Nevertheless, the high crowding and extinction hamper its study, and even its morphology and kinematics are not yet totally clear.}
   {In this work we use NSD red clump stars, whose intrinsic properties are well known, to trace the kinematics of the NSD and to compute the distance and extinction towards the edges of the NSD.} 
    {We used publicly available proper motion and photometric catalogues of the NSD to distinguish red clump stars by using a colour-magnitude diagram. We then applied a Gaussian mixture model to obtain the proper motion distribution, and computed the extinction and distance towards stars with different kinematics.}
   {We obtained that the proper motion distributions contain NSD stars rotating eastwards and westwards, plus some contamination from Galactic bulge/bar stars, in agreement with previous work. We computed the distance and extinction towards the eastward- and westward-moving stars and concluded that the latter are $\sim300$\,pc beyond, indicating a similar structure along and across the line of sight, and  consistent with an axisymmetric structure of the NSD. Moreover, we found that the extinction within the NSD is relatively low and  accounts for less than 10\,\% of the total extinction of the stars belonging to the farthest edge of the NSD.}  
   {}


   \maketitle
%

\section{Introduction}

The nuclear stellar disc (NSD) is a dense stellar structure at the centre of the Galaxy with a radius of $\sim150$\,pc and a scale height of $\sim40$\,pc \citep[e.g.][]{Launhardt:2002nx,gallego-cano2019,Sormani:2020aa,Sormani:2022wv}. It partially overlaps with the central molecular zone \citep[CMZ), a dense accumulation of gas in orbits around the Galactic centre (e.g.][]{Henshaw:2022vl}, and constitutes a distinct structure from the larger Galactic bulge/bar and from the Milky Way's nuclear star cluster located at its centre \citep[e.g.][]{Schultheis:2021wf,Nogueras-Lara:2021wm,Nogueras-Lara:2022tp}. The NSD hosts a total stellar mass of $\sim10^9\,M_\odot$ \citep[e.g.][]{Launhardt:2002nx,Nogueras-Lara:2019ad,Sormani:2022wv}, and is characterised by a predominantly old stellar population ($\sim90$\,\% of the stellar mass is older than 8\,Gyr), but also presents a significant mass of stars ($\sim5$\,\% of the total stellar mass) with an age of $\sim1$\,Gyr \citep{Nogueras-Lara:2019ad}. Moreover, there is evidence of current star formation that makes the NSD the most active star-forming region in the Galaxy when averaged over volume \citep[e.g.][]{Mezger:1996uq,Matsunaga:2011uq,Nogueras-Lara:2019ad,Nogueras-Lara:2022ua}. 

The NSD is located at only 8\,kpc from Earth, which makes it a unique template to understand stellar nuclei and their role in the context of galaxy evolution. Its formation seems to be closely related to the Galactic bulge/bar, which funnelled gas towards the Galactic centre originating the formation of the NSD \citep[e.g.][]{Nogueras-Lara:2019ad}.  Simulations suggest that  the NSD formation is indeed related to the bar, and can be even used to date the Galactic bar \citep[e.g.][]{Baba:2020aa}. Therefore, the NSD is related to the Galaxy as a whole and a detailed analysis of its structure and stellar population is crucial to understanding the origin and properties of the Galactic bulge/bar. However the high extinction and the extreme source crowding towards the Galactic centre challenge the analysis of the NSD stellar population \citep[e.g.][]{Nishiyama:2008qa,Nogueras-Lara:2021wj}. This means that the NSD structure is not clear yet given the complicated determination of the line-of-sight distance towards individual stars.

In this work we use red clump  \citep[RC) stars, red giant stars in their helium core burning sequence, whose intrinsic properties are well known (e.g.][]{Girardi:2016fk} and kinematic data \citep{Libralato:2021td} to compute the line-of-sight distance and extinction towards different regions in the NSD. We obtained the first direct estimate of the NSD depth along the line of sight and conclude that our results are compatible with an axisymmetric morphology of the NSD, with relatively low extinction within it, although based on the present results we cannot rule out that the NSD is barred.

\section{Data}

\subsection{Proper motions}   

\begin{figure*}
   \includegraphics[width=\linewidth]{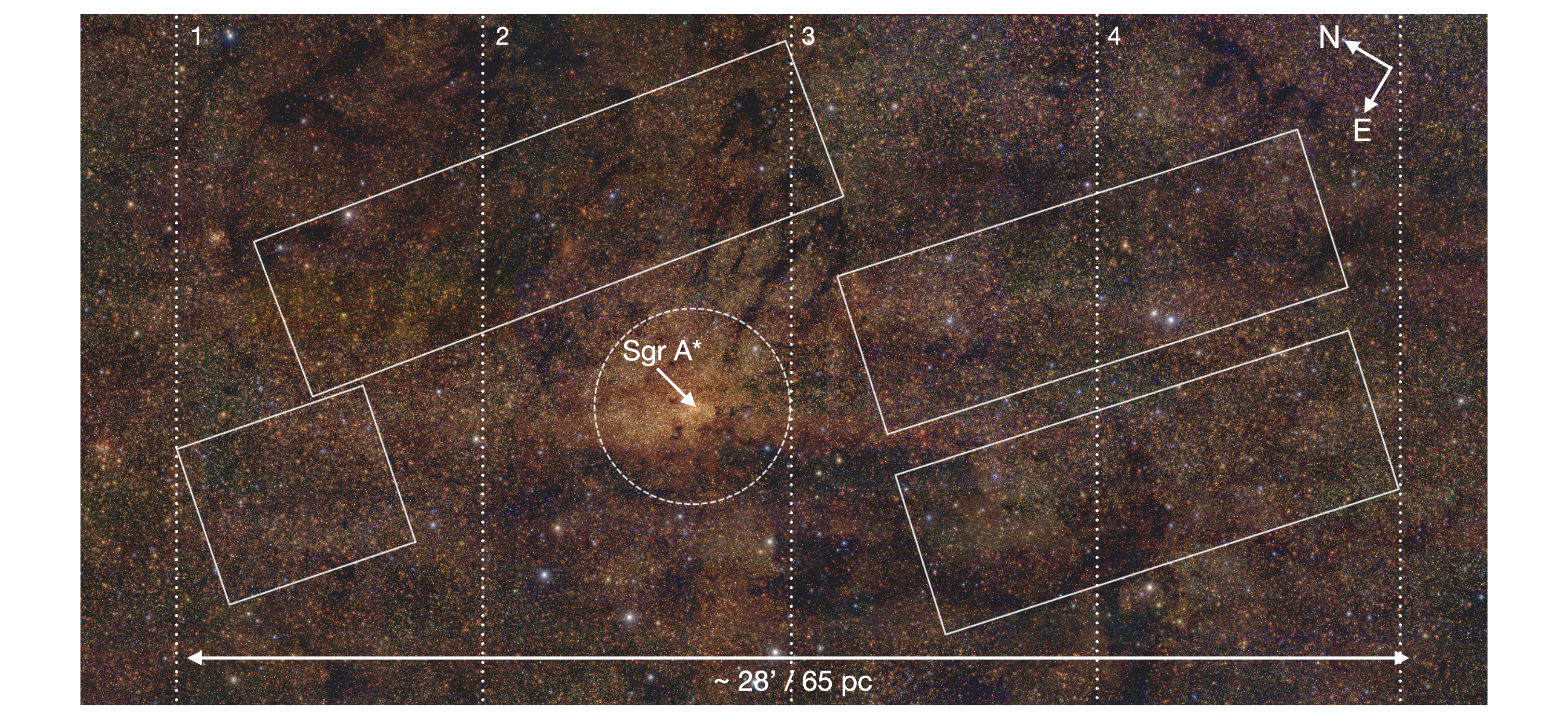}
   \caption{GALACTICNUCLEUS false colour image of the central region of the NSD. The white rectangles indicate the regions covered by the proper motion catalogue. The white dashed circle shows the effective radius of the nuclear star cluster \citep[$\sim 5$\,pc, e.g.][]{gallego-cano2019} with Sagittarius\,A* at its centre. The white dashed lines indicate the four regions used for the distance and extinction analysis in Sect.\,\ref{struct}.}

   \label{GNS}
    \end{figure*}

We used the publicly available NSD proper motion catalogue obtained with the Wide-Field Camera 3 (WFC3/IR, filter F153M) by \citet{Libralato:2021td}. The catalogue contains absolute proper motions for more than 800,000 stars  calibrated with the Gaia Data Release 2 \citep{Gaia-Collaboration:2016uw,Gaia-Collaboration:2018aa}. The proper motions were calculated using point spread function photometry on  images from two different epochs (2012 and 2015). Figure\,\ref{GNS} shows the region covered by the catalogue. For our analysis we removed stars with high uncertainties, and only used proper motions with $\Delta\mu_{ra}$, $\Delta\mu_{dec}<1.5$\,mas/yr. We also applied a rotation to transform the equatorial coordinates into Galactic coordinates to make the analysis of the NSD kinematics easier.

\subsection{Photometry}

We cross-correlated the $H$ and $K_s$ data from the GALACTICNUCLEUS survey \citep{Nogueras-Lara:2018aa,Nogueras-Lara:2019aa} with the proper motion catalogue to build a colour-magnitude diagram (CMD) and identify RC stars for the subsequent analysis. The GALACTICNUCLEUS survey was specially designed to observe the Galactic centre, and contains accurate point spread function photometry for more than three million stars in the NSD and the innermost Galactic bulge/bar. The photometric uncertainties are below 0.05\,mag at $H\sim19$\,mag and $K_s\sim18$\,mag. The zero point systematic uncertainty is below 0.04\,mag.

\begin{figure}
   \includegraphics[width=\linewidth]{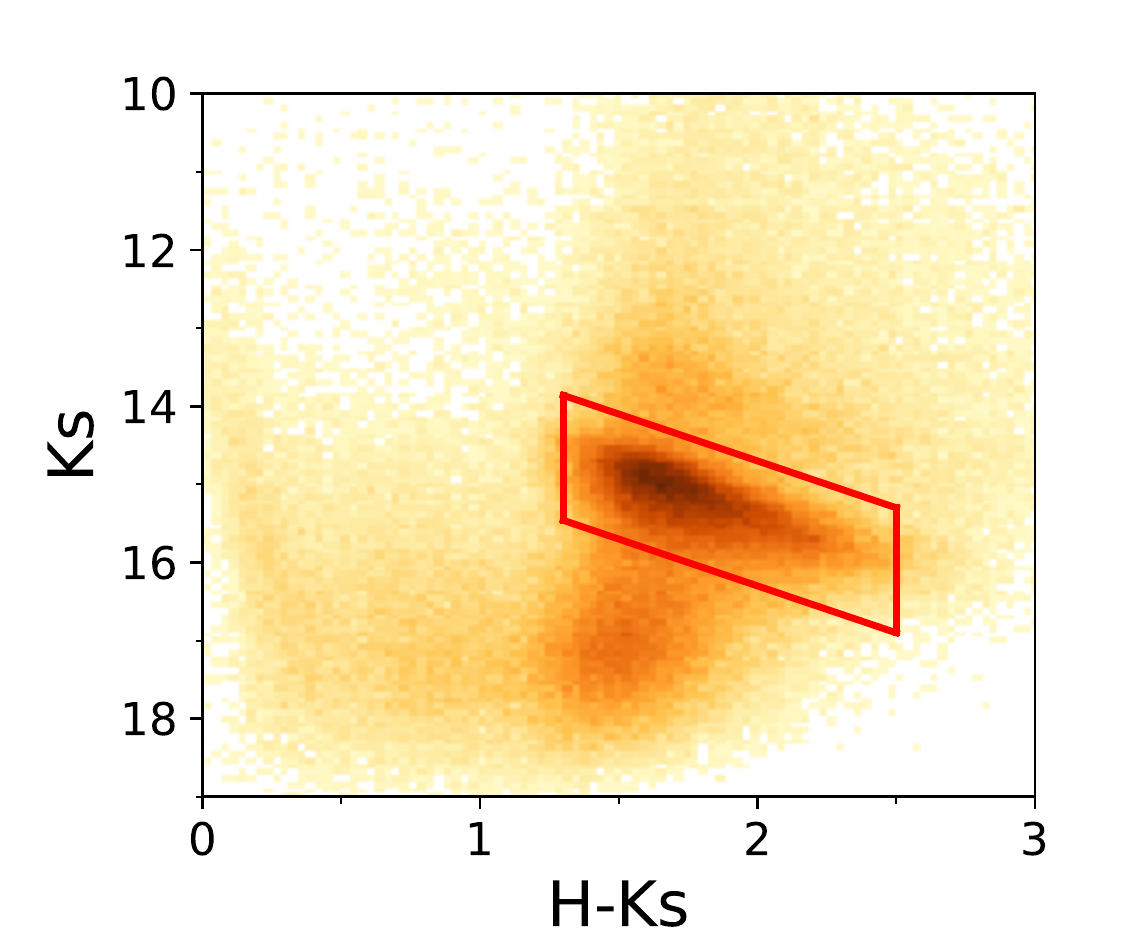}
   \caption{CMD $K_s$ vs $H-K_s$ for common stars between the proper motion catalogue and the GALACTINUCLEUS survey. The red parallelogram shows the selection of RC stars.}

   \label{CMD}
    \end{figure}

Figure\,\ref{CMD} shows the resulting CMD $K_s$ versus $H-K_s$. To select the RC stars from the NSD, we firstly removed foreground stars, which mainly belong to the Galactic disc and the Galactic bulge/bar, by applying a colour cut $H-K_s=1.3$\,mag. This accounts for the significantly different extinction between these Galactic components and the NSD \citep[e.g.][]{Nogueras-Lara:2021uz}. We also assumed a colour cut $H-K_s=2.5$\,mag to avoid regions whose completeness is too low. The red parallelogram in Fig.\,\ref{CMD} shows the resulting selection of RC stars.

\section{NSD kinematics}

\subsection{Proper motion distribution}  
\label{prop_sect}

Recent work \citep{Shahzamanian:2021wu,Martinez-Arranz:2022uf,Nogueras-Lara:2022tp} computed the proper motion distribution of the NSD distinguishing between its parallel and perpendicular components to the Galactic plane ($\mu_l$ and $\mu_b$, respectively). They found that the $\mu_l$ distribution of the NSD presents three stellar components: two of them due to the rotation of the NSD, and the third  probably due to the contamination from the Galactic bulge. On the other hand, the $\mu_b$ distribution shows two components centred around 0 mas/yr that are probably due to the combined contributions from the NSD and the Galactic bulge/bar. 

\begin{figure}
   \includegraphics[width=\linewidth]{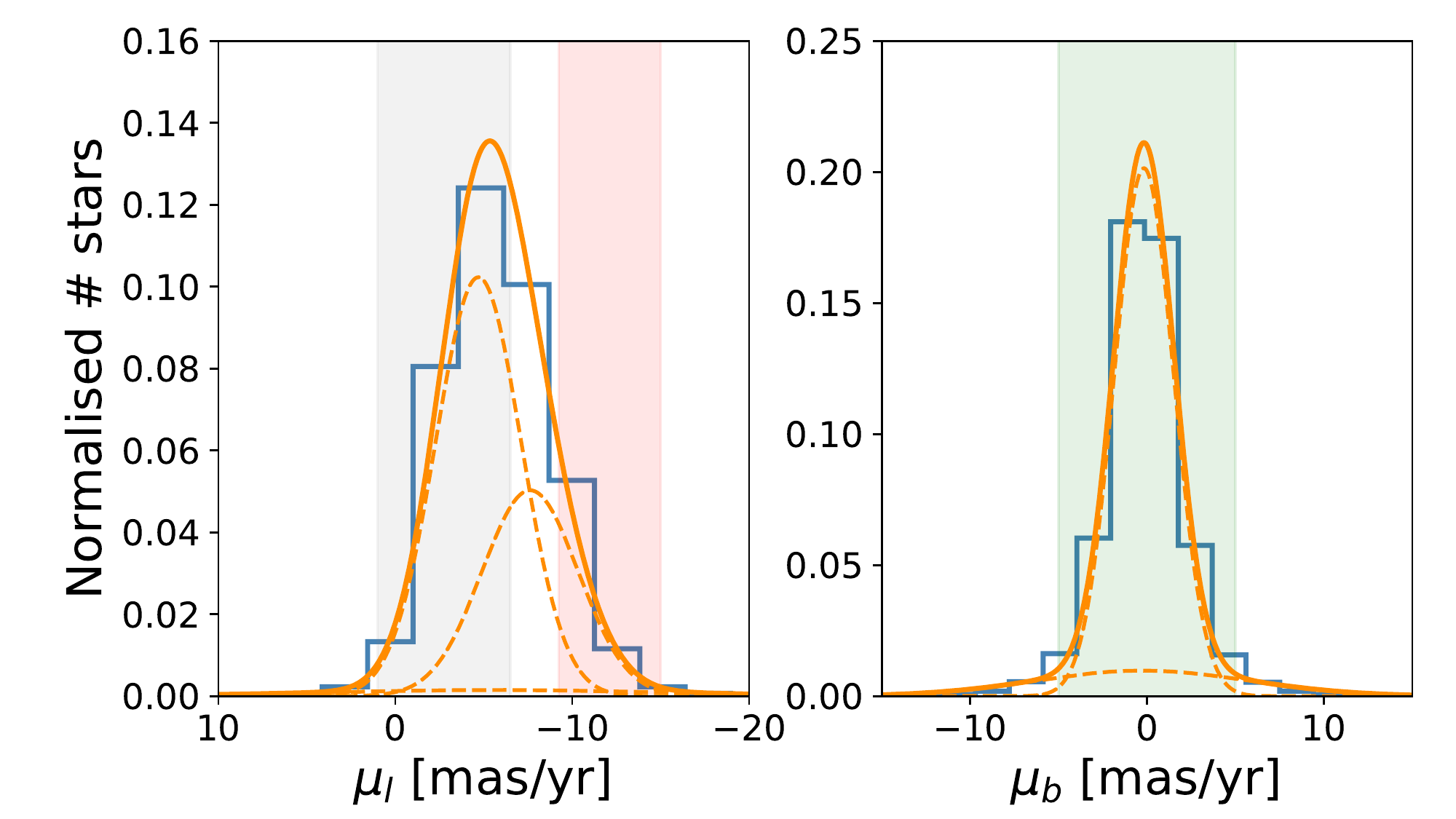}
   \caption{GMM analysis of the proper motion distributions for all the RC stars in the analysed region. The left and right panels show the distribution of the proper motion components parallel ($\mu_l$) and perpendicular ($\mu_b$) to the Galactic plane, respectively. The blue histograms correspond to the data, whereas the orange lines are the total models (solid lines), and each of their Gaussian components (dashed lines). The grey and red shaded areas in the left panel, and the green shaded area in the right panel correspond to the regions analysed in Sect.\,\ref{struct}.}

   \label{proper}
    \end{figure}

Here we use a deeper data set \citep{Libralato:2021td}, which allows us to reach the RC feature and use these stars to trace the stellar kinematics, instead of using a mixture of brighter stars with different stellar types \citep{Shahzamanian:2021wu,Martinez-Arranz:2022uf,Nogueras-Lara:2022tp}. We obtained the distribution of proper motions and applied the SCIKIT-LEARN Python function GaussianMixture \citep[GMM][]{Pedregosa:2011aa} to compute the probability density function that originates the underlying data distribution based on Gaussian models. We tried up to three Gaussians for the $\mu_l$ distribution, and up to two for the $\mu_b$,  in agreement with previous work \citep{Shahzamanian:2021wu,Martinez-Arranz:2022uf,Nogueras-Lara:2022tp}. We computed the Bayesian information criterion \citep[BIC][]{Schwarz:1978aa} and the Akaike information criterion \citep[AIC][]{Akaike:1974aa} to choose the model that best represents the observed distribution. We obtained that the $\mu_l$ distribution is best represented by a three-Gaussian model, whereas the $\mu_b$ distribution is best represented by a two-Gaussian model. Figure\,\ref{proper} and Table\,\ref{GMM} show the obtained results. We estimated the uncertainties of the GMM parameters resorting to Monte Carlo simulations and generating 1000 samples of synthetic data assuming that each data point can vary following a Gaussian distribution with a standard deviation equal to its uncertainty.

\begin{table*}
\caption{Results from the GMM analysis of the proper motion distribution.}
\label{GMM} 
\begin{center}
\def\arraystretch{1.3}
\setlength{\tabcolsep}{3.8pt}        
   
   \begin{tabular}{ccc|ccc}
 &  & \multicolumn{1}{c}{} &  &  & \tabularnewline
\hline 
\hline 
$A_l$ & $\mu_l$ & $\sigma_l$ & $A_b$ & $\mu_b$ & $\sigma_b$\tabularnewline
(norm. units) & (mas/yr) & (mas/yr) & (norm. units) & (mas/yr) & (mas/yr)\tabularnewline
\hline 
0.62 $\pm$ 0.01 &-4.70 $\pm$ 0.03 & 2.48 $\pm$ 0.02 & 0.862 $\pm$ 0.002&  -0.174 $\pm$ 0.003 & 1.81 $\pm$ 0.01 \tabularnewline
0.039 $\pm$ 0.002 &-5.71 $\pm$ 0.04 & 9.48 $\pm$ 0.17 & 0.138 $\pm$ 0.002&  -0.27 $\pm$ 0.01 & 6.23 $\pm$ 0.04 \tabularnewline
0.34 $\pm$ 0.01 &-7.65 $\pm$ 0.03 & 2.79 $\pm$ 0.02 &   &  & \tabularnewline
\hline 
 &  & \multicolumn{1}{c}{} &  &  & \tabularnewline
\end{tabular}

\end{center}
\footnotesize
\textbf{Notes.} $A_i$, $\mu_i$, and $\sigma_i$ respectively indicate the amplitude, the mean value, and the standard deviation of each of the components of the GMM modelling, where the subindex $i$ indicates Galactic longitude ($i=l$) or latitude ($i=b$).

 \end{table*}    

The three-Gaussian distribution found for $\mu_l$ agrees with previous work \citep{Shahzamanian:2021wu,Martinez-Arranz:2022uf,Nogueras-Lara:2022tp} and indicates the presence of two stellar populations moving eastwards and westwards, which we interpret as NSD rotation. Our results are computed using absolute proper motions, which explains why the central peak is not close to zero. We measured a difference of $\sim3$\,mas/yr between the mean values of the $\mu_l$ components moving eastwards and westwards, which is somewhat smaller that the difference obtained in previous work \citep[$\sim4$\,mas/yr, in][]{Shahzamanian:2021wu}. Moreover, the Gaussian model corresponding to the stellar population moving westwards is smaller ($\sim50$\,\% lower) than the model corresponding to the stellar population moving eastwards, whereas they should be similar in size according to previous work \citep{Shahzamanian:2021wu}. We believe that this is because the stellar population moving westwards corresponds to the farthest edge of the NSD \citep{Shahzamanian:2021wu}, and then the extinction towards these stars is higher and their completeness (also affected by the extreme source crowding), is lower than for the eastward-moving stars. The same effect can also cause   the difference between the central peak and the extreme ones to be smaller than measured in previous work. On the other hand, the differences between our results and the work by \citet{Shahzamanian:2021wu} can also be explained because their data set contains the nuclear star cluster, which is not present in our data, whose proper motion distribution is different from the NSD \citep{Nogueras-Lara:2022tp}. Moreover, the discrepancies can also be due to the different techniques used to analyse the data. The work by \citet{Shahzamanian:2021wu} applied a direct fit to the data instead of the GMM modelling that we use in this work. Our approach has the advantage of being independent of the data binning, and informs about the underlying Gaussian model that originates the observed data distribution.

On the other hand, the two-Gaussian model found for the $\mu_b$ distribution agrees well with previous work \citep[e.g.][]{Shahzamanian:2021wu}. We found that the wider Gaussian, probably due to stars from the Galactic bulge/bar, accounts for $\sim15$\,\% of the stars in the distribution. This is smaller than the $\sim 50$\,\% found by \citet{Shahzamanian:2021wu}. We believe that this is due to the smaller region     closer to the Galactic plane  that we consider in this work, which reduces the relative fraction of contaminant stars from the Galactic bulge/bar. Furthermore, our results agree with the expected contamination ($\sim20$\,\%) for the innermost fields of the NSD from Galactic  bulge/bar stars estimated by \citet{Sormani:2022wv} for stars with a colour cut of $H-K_s>1.3$\,mag  (see their Table\,2). The fact that we obtain a contribution of $\sim5$\,\% when considering the central Gaussian of the $\mu_l$ distribution is probably due to the more challenging identification of Galactic bulge/bar stars in this proper motion space in comparison to the $\mu_b$ distribution.  \citet{Shahzamanian:2021wu} also measured a  lower contribution of Galactic bulge/bar stars in the $\mu_l$ distribution ($\sim38\,\%$) in comparison with the $\mu_b$ distribution ($\sim55\,\%$).

\subsection{Testing the asymmetry of the stellar kinematics} 

\begin{figure}
   \includegraphics[width=\linewidth]{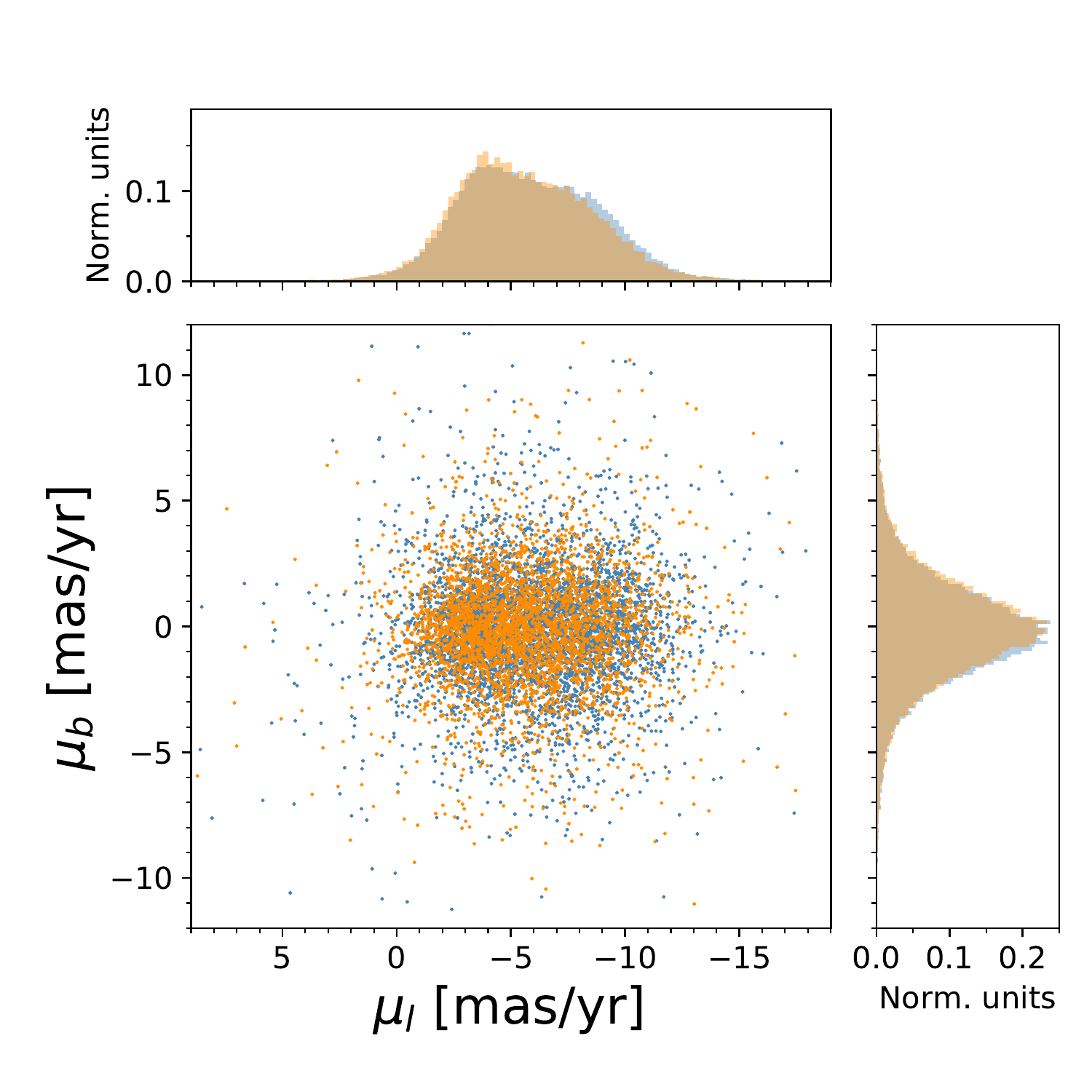}
   \caption{Proper motion distributions of east and west NSD regions. The orange dots and histograms correspond to the proper motions distribution of the region 1+2 in Fig.\,\ref{GNS}, whereas the blue dots and histograms correspond to region 3+4 in Fig.\,\ref{GNS}.}

   \label{half}
    \end{figure}

It is well known that the gas distribution in the innermost 300\,pc of the Galaxy is strongly asymmetric \citep[e.g.][]{Bally:1988vf,Sormani:2018aa}. We tested whether the stellar kinematics is also different when considering different regions of the NSD in the following way. We divided our proper motion data set into two different regions (1+2 and 3+4 in Fig.\,\ref{GNS}), corresponding to different longitudes of the NSD. Figure\,\ref{half} shows the proper motion distribution for each of the regions. We conclude that the $\mu_l$ and $\mu_b$ distributions are similar for both regions of the NSD, so there is no remarkable difference between the stellar kinematics depending on the observed regions.

\section{NSD structure along the line of sight}
\label{struct}

Red clump stars have well-characterised intrinsic properties \citep[e.g.][]{Girardi:2016fk}, such as their absolute magnitude  \citep[M$_{K}=-1.54\pm0.04$\,mag, ][]{Groenewegen:2008by} and their intrinsic colour \citep[$(H-K_s)_0=0.10\pm0.01$\,mag,][]{Nogueras-Lara:2021wj}. They are thus adequate to estimate distance and extinction towards stellar populations \citep[e.g.][]{Nishiyama:2006ai,Cabrera-Lavers:2007fk,Nogueras-Lara:2021wj,Nogueras-Lara:2021uq}. 

\subsection{Stellar sample}

Using the proper motion distribution derived in Sect.\,\ref{prop_sect}, we obtained a clean sample of RC stars belonging to the eastward- and westward-moving stars from the NSD; they  trace the stellar population from the closest edge (eastward-moving stars) and the farthest one (westward-moving stars) of the NSD. For this, we firstly removed a significant fraction of the contaminant stars from the Galactic bulge/bar by choosing stars with $|\mu_b|<5$\,mas/yr (green shaded region in Fig.\,\ref{proper}), which reduces their contamination below 10\,\%. We then computed the probability of a given star to be part of each of the three-Gaussian models that originate the $\mu_l$ distribution. We defined two regions in the $\mu_l$ distribution that are dominated (i.e. $\sim 80\,\%$ of the stars) by stars moving eastwards (grey shaded area in Fig.\,\ref{proper}), and by stars moving westwards (red shaded area in Fig.\,\ref{proper}). These regions are not symmetric and have different sizes because the corresponding underlying Gaussian distributions are also not symmetric due to the lower completeness of the stellar sample moving westwards (see Sect.\,\ref{prop_sect}). To analyse the behaviour of the NSD population for different lines of sight, we divided the stellar samples into four different regions, each with a similar number of stars, as shown in Fig.\,\ref{GNS}. 

\subsection{Extinction}

We computed extinction to de-redden the stars in each region using a star-by-star approach applying the equation \citep[e.g.][]{Nogueras-Lara:2021wm,Nogueras-Lara:2021uq}

\begin{equation}
\label{eq}
A_{K_s}=\frac{H-K_s-(H-K_s)_0}{A_{H}/A_{K_s} -1}\hspace{0.25cm},
\end{equation}

\noindent where $A_{Ks}$ is the $K_s$ extinction, $H$ and $K_s$ indicate the photometric data, $(H-K_s)_0 = 0.10\pm0.01$ is the intrinsic colour \citep{Nogueras-Lara:2021wj}, and $A_{H}/A_{K_s}=1.84\pm0.03$ is the extinction curve \citep{Nogueras-Lara:2020aa}. Table\,\ref{results_distance} shows the mean extinction values obtained by applying a 3$\sigma$ clipping algorithm to remove outliers. The uncertainties were obtained quadratically propagating the uncertainties of the quantities involved in the calculation. Moreover, we also computed the relative difference in extinction between the stellar populations moving eastwards and westwards. In this case the uncertainties refer to the standard error of the distribution and to the uncertainty associated with the used extinction curve. Other systematic uncertainties from the quantities involved in Eq.\,\ref{eq} do not affect the difference between the extinctions towards each of the stellar populations.  

We obtained that the extinction is larger for the westward-moving stellar population, which is in agreement with this population belonging to the farthest edge of the NSD. Comparing the extinctions between the eastward- and the westward-moving stars, we found that $<10\,\%$ of the total extinction occurs within the NSD. This suggests that the majority of gas and dust causing the extinction in the NSD is not inside it, which may indicate that most of the gas in the CMZ surrounds the NSD \citep[e.g.][]{Henshaw:2022vl}. This can be related with the inside-out formation of nuclear stellar discs suggested by \citet{Bittner:2020aa} based on observations of external galaxies. In this picture, the CMZ would be the star-forming outer edge of the NSD.

\begin{figure}
   \includegraphics[width=\linewidth]{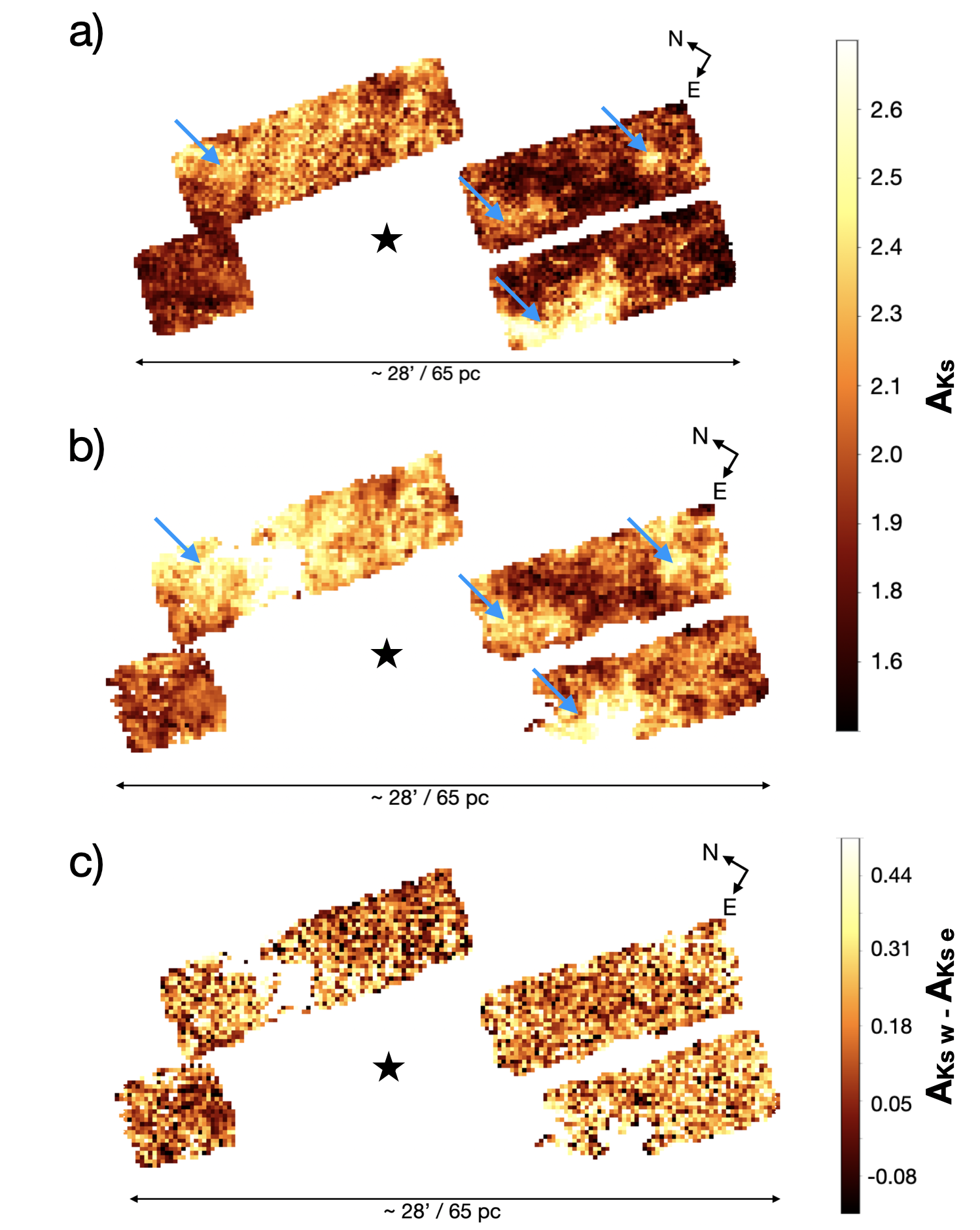}
   \caption{Extinction maps towards the analysed regions (panels a and b, built using eastward- and westward-moving RC stars, respectively), and their difference (panel c). The black star indicates the position of Sagittarius\,A*. The blue arrows in panels a and b correspond to regions with similar extinction structures in both maps. The white pixels indicate that there is not any extinction value associated, due to the low number of reference stars.}

   \label{extinction_maps}
    \end{figure}

An alternative explanation for the relatively low extinction within the NSD might be an extremely clumpy distribution of the dust and gas inside the NSD that would leave the majority of stars from the farthest edge of the NSD relatively unobscured. In this way, some regions of the NSD could contain a large quantity of dust and gas. To check this scenario, we produced extinction maps for the closest and the farthest NSD edges using the analysed RC stars. We followed the process described in \citet{Nogueras-Lara:2021wj}, and chose a pixel size of $10''$ for the extinction maps. We computed an extinction value for each pixel using the five closest RC stars in a radius of $15''$ (if fewer than five stars are present within the reference radius for a given pixel, we did not compute an associated value). Figure\,\ref{extinction_maps} shows the obtained extinction maps. We also computed the difference between the two extinction maps (Fig.\,\ref{extinction_maps}c). The main extinction structures appearing in both extinction layers are not present in the difference map, and   thus affect the two stellar components in the same way,  probably being due to the presence of dust and gas along the line of sight towards the Galactic centre, and likely belonging to the closest edge of the CMZ. We computed the mean value of the difference map and its standard deviation, $0.18\pm0.16$\,mag, and obtained that less than 0.5\,\% of the pixels have an extinction $>3\sigma$ away from the mean (corresponding to regions with a potential clumpy distribution of dust and gas), and thus the extinction map is quite homogeneous. Even for the regions 3$\sigma$ away from the mean value, the extinction within the NSD is only $\lesssim30\,\%$ of the total extinction along the line of sight.

    \begin{table*}
\caption{Results from the analysis of the de-reddened RC luminosity functions.}
\label{results_distance} 
\begin{center}
\def\arraystretch{1.3}
\setlength{\tabcolsep}{3.8pt}

    \begin{tabular}{cccccccc}
 &  &  &  &  &  &  & \tabularnewline
\hline 
\hline 
Region & \#eastward stars & \#westward stars & $A_{eastwards}$ & $A_{westwards}$ & $\Delta A_{K_s}$ & $\Delta K_{s}$ & $\Delta d$\tabularnewline
 & (\# of stars) & (\# of stars) & (mag) & (mag) & (mag) & (mag) & (pc)\tabularnewline
\hline 
1 & 12937 & 1771 & 1.97 $\pm$ 0.11 & 2.17 $\pm$ 0.10 & 0.20 $\pm$ 0.01 & 0.11 $\pm$ 0.02 & 355 $\pm$ 80\tabularnewline
2 & 8275 & 1912 & 2.13 $\pm$ 0.11 & 2.29 $\pm$ 0.11 & 0.16 $\pm$ 0.01 & 0.08 $\pm$ 0.02 & 254 $\pm$ 84\tabularnewline
3 & 12677 & 2492 & 1.93 $\pm$ 0.10 & 2.08 $\pm$ 0.10 & 0.15 $\pm$ 0.01 & 0.11 $\pm$ 0.02 & 380 $\pm$ 72\tabularnewline
4 & 13982 & 3534 & 1.88 $\pm$ 0.10 & 2.09 $\pm$ 0.10 & 0.21 $\pm$ 0.01 & 0.10 $\pm$ 0.02 & 330 $\pm$ 64\tabularnewline
\hline 
 &  &  &  &  &  &  & \tabularnewline
\end{tabular}

\end{center}
\footnotesize
\textbf{Notes.} Region indicates the corresponding field   indicated in Fig.\,\ref{GNS}. $A_{i}$ indicates the absolute extinction for each of the $i$ stellar populations. $\Delta A_{K_s}$ is the relative extinction between the analysed stellar populations. $\Delta K_{s}$ corresponds to the de-reddened magnitude difference between the stellar populations. $\Delta d$ indicates the relative distance between the stellar populations.

 \end{table*}

\subsection{Relative distance}

\begin{figure*}
   \includegraphics[width=\linewidth]{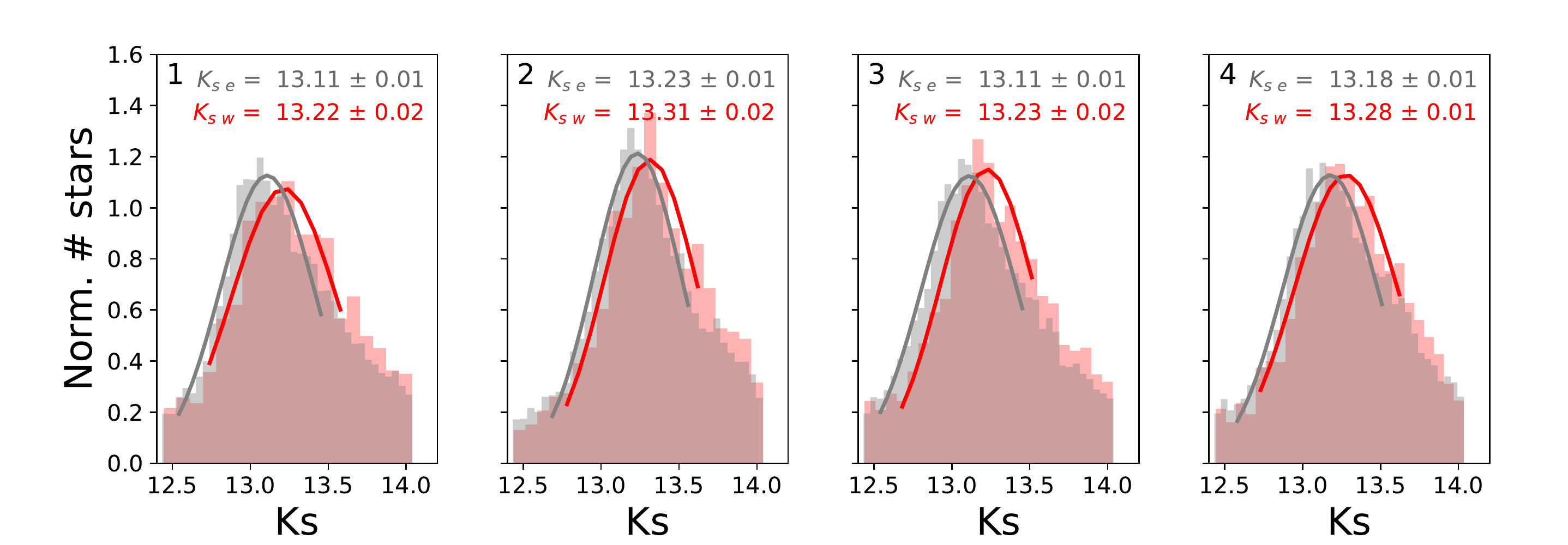}
   \caption{De-reddened RC $K_s$ luminosity function for different NSD regions. The  histograms, Gaussian fits, and numbers  in each panel refer to the eastward-moving stellar population (in grey) and the  westward-moving stars (in red). The numbers in each panel indicate the associated region in Fig.\,\ref{GNS}. The $K_{s\ e}$ and $K_{s\ w}$ values indicate the mean value of the Gaussian fits and its associated uncertainty for the eastward- and westward-moving stars.}

   \label{distances}
    \end{figure*}

We used the previously computed extinctions to build de-reddened $K_s$ luminosity functions for the RC stars of the eastward- and westward-moving stellar populations for each of the four NSD regions analysed. Given the stellar population present in the NSD, we expected to find two peaks in the RC bump. The brightest peak, around $K_s\sim 13$\,mag, is caused by a predominantly old stellar population (more than 80\,\% of the stellar mass is older than 8\,Gyr). The secondary peak is   around $\sim 0.5$\,mag fainter than the main peak, and it is produced by RC stars with an age of around 1\,Gyr \citep{Nogueras-Lara:2021wj}. We fitted the underlying distributions for each region and stellar population with different kinematics using a Gaussian model. To account for the possible influence of the double peak in the luminosity functions, we computed the maximum ($max$) of each distribution and restricted the analysis to stars with $K_s\in[max-0.55, max+0.45]$\,mag. In this way we guarantee that the fit is consistent and that  the possible influence of the secondary peak is similar for all the regions and stellar groups with different kinematics. Figure\,\ref{distances} shows the results. We obtained that the mean magnitude of the RC bump is always brighter for the eastward-moving stellar population in comparison with the westward-moving one. Our results agree with a different distance between eastward- and westward-moving stars, as suggested by \citet{Shahzamanian:2021wu}.

We computed the relative distance between the eastward- and the westward-moving stellar populations for each of the regions, by applying the equation

\begin{equation}
r_{d}  = 10^{\Delta \mu/5} = 10^{\Delta K_{s}/5} \hspace{0.25cm},
\end{equation}

\noindent where $r_{d}$ is the ratio of the distances towards the two stellar populations, $\Delta \mu$ is the difference in their distance modulus, and $\Delta K_{s}$ is their de-reddened magnitude difference. To compute the distance difference between the two stellar populations, we assumed a Galactic centre distance of 8.1\,kpc \citep{Gravity-Collaboration:2018aa}, and considered that the closest edge of the NSD is 150\,pc closer than that \citep[e.g.][]{Launhardt:2002nx,gallego-cano2019,Sormani:2020aa,Sormani:2022wv}. In this way we placed the eastward-moving stellar population at a distance of 7.95\,kpc and used the ratio $r_d$ to estimate the relative distance between the components. Table\,\ref{results_distance} shows the obtained results. The associated uncertainties were computed considering the uncertainty of the relative distance modulus ($\Delta \mu$), which includes the uncertainty of the Gaussian fits, and also the relative uncertainty due to the de-reddening process. Given that we are interested in the relative distance between the stellar populations, the systematic uncertainties related to the photometric zero point, the absolute distance used to scale the values, the extinction curve, and the intrinsic colour do not affect the computed uncertainty.

We obtained that the distance between the two stellar populations is similar and compatible within the uncertainties for all the analysed regions. Given that the 2D projected radius of the NSD was estimated to be $\sim150$\,pc \citep[e.g.][]{Launhardt:2002nx,gallego-cano2019,Sormani:2020aa,Sormani:2022wv}, our results suggest a similar structure along and across the line of sight. This, in combination with  the symmetric proper motion distribution in the analysed region (see Fig.\,\ref{half}) and the rotation pattern of the NSD obtained from line-of-sight velocities using APOGEE data \citep{2015ApJ...812L..21S}, is compatible with an axisymmetric geometry of the NSD, although a nuclear bar as found in external galaxies \citep{Bittner:2021wu} cannot be excluded.

We computed a mean relative distance value between the two stellar populations averaging over the results for each of the four analysed regions. We obtained $\bar{d} = 330\pm50$\,pc, where the uncertainty corresponds to the standard deviation of the distribution. This value is compatible with the previously estimated NSD radius of about 150\,pc \citep[e.g.][]{Launhardt:2002nx,Schodel:2014fk,gallego-cano2019,Sormani:2020aa,Sormani:2022wv}. Nevertheless, the obtained value is instead a lower limit of the NSD diameter because it corresponds with the relative distance between the stellar populations considered in this work, and might be lower than the actual diameter because our sample of RC stars is probably not complete at its faint end. In any case, the depth along the line of sight is comparable to the extension across the line of sight, consistent with an axisymmetric structure of the NSD. On the other hand, this analysis does not imply that the NSD is necessarily a ring because we did not consider the transition region where stars from the eastward- and westward-moving populations are mixed (white space between the red and grey shaded regions in Fig.\,\ref{proper}), and might correspond to intermediate distances, building a disc-like geometry or even a nuclear bar.

\section{Conclusion}

In this letter we studied the kinematics of RC stars from the NSD. We analysed their proper motion components parallel ($\mu_l$) and perpendicular ($\mu_b$) to the Galactic plane by applying a GMM modelling. We obtained that the $\mu_l$ component is best described by a three-Gaussian model, which  we interpret as a result of the rotation of the NSD in combination with some residual contamination from Galactic bulge/bar stars. The $\mu_b$ distribution is described by a two-Gaussian model that accounts for the NSD stars and also the contaminant stars from the Galactic bulge/bar. Our results are consistent with previous work \citep{Shahzamanian:2021wu,Martinez-Arranz:2022uf,Nogueras-Lara:2022tp}, with the exception of a somewhat smaller contribution from Galactic bulge/bar stars. We believe that this is due to the region used in this work,  smaller and closer to the Galactic plane, which reduces the contamination from the Galactic bulge/bar. Moreover, we divided the data set into two regions with different Galactic longitudes and checked that the proper motion distributions are similar, pointing towards a symmetric NSD structure.

We also created a clean sample of RC stars belonging to the eastward- and westward-moving stellar populations, and computed their relative distance and extinction towards four subregions covering different Galactic longitudes. We obtained that the relative distance between the two stellar populations is constant within the uncertainties. These results, in combination with previous measurements of the NSD extension across the line of sight \citep[e.g.][]{Launhardt:2002nx, gallego-cano2019} and the symmetric distribution of the proper motions, are consistent with the NSD being axisymmetric, although the presence of a nuclear bar cannot be ruled out with the current measurements. Moreover, we find that the extinction within the NSD is relatively small in comparison with the overall extinction of the NSD stars. Our findings suggest that the most of the extinction in the Galactic centre is due to the presence of the CMZ that surrounds the NSD, being the amount of gas and dust inside the NSD relatively small.

  \begin{acknowledgements}
F. N.-L. gratefully acknowledges the sponsorship provided by the Federal Ministry for Education and Research of Germany through the Alexander von Humboldt Foundation. This work is based on observations made with ESO Telescopes at the La Silla Paranal Observatory under program ID195.B-0283. We thank the anonymous referee for comments and suggestions.

\end{acknowledgements}

\bibliography{../../BibGC.bib}

\begin{thebibliography}{39}
\expandafter\ifx\csname natexlab\endcsname\relax\def\natexlab#1{#1}\fi

\bibitem[{Akaike(1974)}]{Akaike:1974aa}
Akaike, H. 1974, Automatic Control, IEEE Transactions on, 19, 716

\bibitem[{{Baba} \& {Kawata}(2020)}]{Baba:2020aa}
{Baba}, J. \& {Kawata}, D. 2020, \mnras, 492, 4500

\bibitem[{{Bally} {et~al.}(1988){Bally}, {Stark}, {Wilson}, \&
  {Henkel}}]{Bally:1988vf}
{Bally}, J., {Stark}, A.~A., {Wilson}, R.~W., \& {Henkel}, C. 1988, \apj, 324,
  223

\bibitem[{{Bittner} {et~al.}(2021){Bittner}, {de Lorenzo-C{\'a}ceres},
  {Gadotti}, {S{\'a}nchez-Bl{\'a}zquez}, {Neumann}, {Coelho},
  {Falc{\'o}n-Barroso}, {Fragkoudi}, {Kim}, {Mart{\'\i}n-Navarro},
  {M{\'e}ndez-Abreu}, {P{\'e}rez}, {Querejeta}, \& {van de
  Ven}}]{Bittner:2021wu}
{Bittner}, A., {de Lorenzo-C{\'a}ceres}, A., {Gadotti}, D.~A., {et~al.} 2021,
  \aap, 646, A42

\bibitem[{{Bittner} {et~al.}(2020){Bittner}, {S{\'a}nchez-Bl{\'a}zquez},
  {Gadotti}, {Neumann}, {Fragkoudi}, {Coelho}, {de Lorenzo-C{\'a}ceres},
  {Falc{\'o}n-Barroso}, {Kim}, {Leaman}, {Mart{\'\i}n-Navarro},
  {M{\'e}ndez-Abreu}, {P{\'e}rez}, {Querejeta}, {Seidel}, \& {van de
  Ven}}]{Bittner:2020aa}
{Bittner}, A., {S{\'a}nchez-Bl{\'a}zquez}, P., {Gadotti}, D.~A., {et~al.} 2020,
  \aap, 643, A65

\bibitem[{{Cabrera-Lavers} {et~al.}(2007){Cabrera-Lavers}, {Hammersley},
  {Gonz{\'a}lez-Fern{\'a}ndez}, {L{\'o}pez-Corredoira}, {Garz{\'o}n}, \&
  {Mahoney}}]{Cabrera-Lavers:2007fk}
{Cabrera-Lavers}, A., {Hammersley}, P.~L., {Gonz{\'a}lez-Fern{\'a}ndez}, C.,
  {et~al.} 2007, \aap, 465, 825

\bibitem[{{Gaia Collaboration} {et~al.}(2018){Gaia Collaboration}, {Brown},
  {Vallenari}, {Prusti}, {de Bruijne}, {Babusiaux}, {Bailer-Jones}, {Biermann},
  {Evans}, {Eyer}, {Jansen}, {Jordi}, {Klioner}, {Lammers}, {Lindegren},
  {Luri}, {Mignard}, {Panem}, {Pourbaix}, {Randich}, {Sartoretti}, {Siddiqui},
  {Soubiran}, {van Leeuwen}, {Walton}, {Arenou}, {Bastian}, {Cropper},
  {Drimmel}, {Katz}, {Lattanzi}, {Bakker}, {Cacciari}, {Casta{\~n}eda},
  {Chaoul}, {Cheek}, {De Angeli}, {Fabricius}, {Guerra}, {Holl}, {Masana},
  {Messineo}, {Mowlavi}, {Nienartowicz}, {Panuzzo}, {Portell}, {Riello},
  {Seabroke}, {Tanga}, {Th{\'e}venin}, {Gracia-Abril}, {Comoretto},
  {Garcia-Reinaldos}, {Teyssier}, {Altmann}, {Andrae}, {Audard},
  {Bellas-Velidis}, {Benson}, {Berthier}, {Blomme}, {Burgess}, {Busso},
  {Carry}, {Cellino}, {Clementini}, {Clotet}, {Creevey}, {Davidson}, {De
  Ridder}, {Delchambre}, {Dell'Oro}, {Ducourant},
  {Fern{\'a}ndez-Hern{\'a}ndez}, {Fouesneau}, {Fr{\'e}mat}, {Galluccio},
  {Garc{\'\i}a-Torres}, {Gonz{\'a}lez-N{\'u}{\~n}ez}, {Gonz{\'a}lez-Vidal},
  {Gosset}, {Guy}, {Halbwachs}, {Hambly}, {Harrison}, {Hern{\'a}ndez},
  {Hestroffer}, {Hodgkin}, {Hutton}, {Jasniewicz}, {Jean-Antoine-Piccolo},
  {Jordan}, {Korn}, {Krone-Martins}, {Lanzafame}, {Lebzelter}, {L{\"o}ffler},
  {Manteiga}, {Marrese}, {Mart{\'\i}n-Fleitas}, {Moitinho}, {Mora}, {Muinonen},
  {Osinde}, {Pancino}, {Pauwels}, {Petit}, {Recio-Blanco}, {Richards},
  {Rimoldini}, {Robin}, {Sarro}, {Siopis}, {Smith}, {Sozzetti}, {S{\"u}veges},
  {Torra}, {van Reeven}, {Abbas}, {Abreu Aramburu}, {Accart}, {Aerts},
  {Altavilla}, {{\'A}lvarez}, {Alvarez}, {Alves}, {Anderson}, {Andrei},
  {Anglada Varela}, {Antiche}, {Antoja}, {Arcay}, {Astraatmadja}, {Bach},
  {Baker}, {Balaguer-N{\'u}{\~n}ez}, {Balm}, {Barache}, {Barata}, {Barbato},
  {Barblan}, {Barklem}, {Barrado}, {Barros}, {Barstow}, {Bartholom{\'e}
  Mu{\~n}oz}, {Bassilana}, {Becciani}, {Bellazzini}, {Berihuete}, {Bertone},
  {Bianchi}, {Bienaym{\'e}}, {Blanco-Cuaresma}, {Boch}, {Boeche}, {Bombrun},
  {Borrachero}, {Bossini}, {Bouquillon}, {Bourda}, {Bragaglia}, {Bramante},
  {Breddels}, {Bressan}, {Brouillet}, {Br{\"u}semeister}, {Brugaletta},
  {Bucciarelli}, {Burlacu}, {Busonero}, {Butkevich}, {Buzzi}, {Caffau},
  {Cancelliere}, {Cannizzaro}, {Cantat-Gaudin}, {Carballo}, {Carlucci},
  {Carrasco}, {Casamiquela}, {Castellani}, {Castro-Ginard}, {Charlot},
  {Chemin}, {Chiavassa}, {Cocozza}, {Costigan}, {Cowell}, {Crifo}, {Crosta},
  {Crowley}, {Cuypers}, {Dafonte}, {Damerdji}, {Dapergolas}, {David}, {David},
  {de Laverny}, {De Luise}, {De March}, {de Martino}, {de Souza}, {de Torres},
  {Debosscher}, {del Pozo}, {Delbo}, {Delgado}, {Delgado}, {Di Matteo},
  {Diakite}, {Diener}, {Distefano}, {Dolding}, {Drazinos}, {Dur{\'a}n},
  {Edvardsson}, {Enke}, {Eriksson}, {Esquej}, {Eynard Bontemps}, {Fabre},
  {Fabrizio}, {Faigler}, {Falc{\~a}o}, {Farr{\`a}s Casas}, {Federici},
  {Fedorets}, {Fernique}, {Figueras}, {Filippi}, {Findeisen}, {Fonti},
  {Fraile}, {Fraser}, {Fr{\'e}zouls}, {Gai}, {Galleti}, {Garabato},
  {Garc{\'\i}a-Sedano}, {Garofalo}, {Garralda}, {Gavel}, {Gavras}, {Gerssen},
  {Geyer}, {Giacobbe}, {Gilmore}, {Girona}, {Giuffrida}, {Glass}, {Gomes},
  {Granvik}, {Gueguen}, {Guerrier}, {Guiraud}, {Guti{\'e}rrez-S{\'a}nchez},
  {Haigron}, {Hatzidimitriou}, {Hauser}, {Haywood}, {Heiter}, {Helmi}, {Heu},
  {Hilger}, {Hobbs}, {Hofmann}, {Holland}, {Huckle}, {Hypki}, {Icardi},
  {Jan{\ss}en}, {Jevardat de Fombelle}, {Jonker}, {Juh{\'a}sz}, {Julbe},
  {Karampelas}, {Kewley}, {Klar}, {Kochoska}, {Kohley}, {Kolenberg},
  {Kontizas}, {Kontizas}, {Koposov}, {Kordopatis}, {Kostrzewa-Rutkowska},
  {Koubsky}, {Lambert}, {Lanza}, {Lasne}, {Lavigne}, {Le Fustec}, {Le
  Poncin-Lafitte}, {Lebreton}, {Leccia}, {Leclerc}, {Lecoeur-Taibi},
  {Lenhardt}, {Leroux}, {Liao}, {Licata}, {Lindstr{\o}m}, {Lister}, {Livanou},
  {Lobel}, {L{\'o}pez}, {Managau}, {Mann}, {Mantelet}, {Marchal}, {Marchant},
  {Marconi}, {Marinoni}, {Marschalk{\'o}}, {Marshall}, {Martino}, {Marton},
  {Mary}, {Massari}, {Matijevi{\v{c}}}, {Mazeh}, {McMillan}, {Messina},
  {Michalik}, {Millar}, {Molina}, {Molinaro}, {Moln{\'a}r}, {Montegriffo},
  {Mor}, {Morbidelli}, {Morel}, {Morris}, {Mulone}, {Muraveva}, {Musella},
  {Nelemans}, {Nicastro}, {Noval}, {O'Mullane}, {Ord{\'e}novic},
  {Ord{\'o}{\~n}ez-Blanco}, {Osborne}, {Pagani}, {Pagano}, {Pailler},
  {Palacin}, {Palaversa}, {Panahi}, {Pawlak}, {Piersimoni}, {Pineau}, {Plachy},
  {Plum}, {Poggio}, {Poujoulet}, {Pr{\v{s}}a}, {Pulone}, {Racero}, {Ragaini},
  {Rambaux}, {Ramos-Lerate}, {Regibo}, {Reyl{\'e}}, {Riclet}, {Ripepi}, {Riva},
  {Rivard}, {Rixon}, {Roegiers}, {Roelens}, {Romero-G{\'o}mez}, {Rowell},
  {Royer}, {Ruiz-Dern}, {Sadowski}, {Sagrist{\`a} Sell{\'e}s}, {Sahlmann},
  {Salgado}, {Salguero}, {Sanna}, {Santana-Ros}, {Sarasso}, {Savietto},
  {Schultheis}, {Sciacca}, {Segol}, {Segovia}, {S{\'e}gransan}, {Shih},
  {Siltala}, {Silva}, {Smart}, {Smith}, {Solano}, {Solitro}, {Sordo}, {Soria
  Nieto}, {Souchay}, {Spagna}, {Spoto}, {Stampa}, {Steele},
  {Steidelm{\"u}ller}, {Stephenson}, {Stoev}, {Suess}, {Surdej}, {Szabados},
  {Szegedi-Elek}, {Tapiador}, {Taris}, {Tauran}, {Taylor}, {Teixeira},
  {Terrett}, {Teyssand ier}, {Thuillot}, {Titarenko}, {Torra Clotet}, {Turon},
  {Ulla}, {Utrilla}, {Uzzi}, {Vaillant}, {Valentini}, {Valette}, {van Elteren},
  {Van Hemelryck}, {van Leeuwen}, {Vaschetto}, {Vecchiato}, {Veljanoski},
  {Viala}, {Vicente}, {Vogt}, {von Essen}, {Voss}, {Votruba}, {Voutsinas},
  {Walmsley}, {Weiler}, {Wertz}, {Wevers}, {Wyrzykowski}, {Yoldas},
  {{\v{Z}}erjal}, {Ziaeepour}, {Zorec}, {Zschocke}, {Zucker}, {Zurbach}, \&
  {Zwitter}}]{Gaia-Collaboration:2018aa}
{Gaia Collaboration}, {Brown}, A.~G.~A., {Vallenari}, A., {et~al.} 2018, \aap,
  616, A1

\bibitem[{{Gaia Collaboration} {et~al.}(2016){Gaia Collaboration}, {Prusti},
  {de Bruijne}, {Brown}, {Vallenari}, {Babusiaux}, {Bailer-Jones}, {Bastian},
  {Biermann}, {Evans}, {Eyer}, {Jansen}, {Jordi}, {Klioner}, {Lammers},
  {Lindegren}, {Luri}, {Mignard}, {Milligan}, {Panem}, {Poinsignon},
  {Pourbaix}, {Randich}, {Sarri}, {Sartoretti}, {Siddiqui}, {Soubiran},
  {Valette}, {van Leeuwen}, {Walton}, {Aerts}, {Arenou}, {Cropper}, {Drimmel},
  {H{\o}g}, {Katz}, {Lattanzi}, {O'Mullane}, {Grebel}, {Holland}, {Huc},
  {Passot}, {Bramante}, {Cacciari}, {Casta{\~n}eda}, {Chaoul}, {Cheek}, {De
  Angeli}, {Fabricius}, {Guerra}, {Hern{\'a}ndez}, {Jean-Antoine-Piccolo},
  {Masana}, {Messineo}, {Mowlavi}, {Nienartowicz}, {Ord{\'o}{\~n}ez-Blanco},
  {Panuzzo}, {Portell}, {Richards}, {Riello}, {Seabroke}, {Tanga},
  {Th{\'e}venin}, {Torra}, {Els}, {Gracia-Abril}, {Comoretto},
  {Garcia-Reinaldos}, {Lock}, {Mercier}, {Altmann}, {Andrae}, {Astraatmadja},
  {Bellas-Velidis}, {Benson}, {Berthier}, {Blomme}, {Busso}, {Carry},
  {Cellino}, {Clementini}, {Cowell}, {Creevey}, {Cuypers}, {Davidson}, {De
  Ridder}, {de Torres}, {Delchambre}, {Dell'Oro}, {Ducourant}, {Fr{\'e}mat},
  {Garc{\'\i}a-Torres}, {Gosset}, {Halbwachs}, {Hambly}, {Harrison}, {Hauser},
  {Hestroffer}, {Hodgkin}, {Huckle}, {Hutton}, {Jasniewicz}, {Jordan},
  {Kontizas}, {Korn}, {Lanzafame}, {Manteiga}, {Moitinho}, {Muinonen},
  {Osinde}, {Pancino}, {Pauwels}, {Petit}, {Recio-Blanco}, {Robin}, {Sarro},
  {Siopis}, {Smith}, {Smith}, {Sozzetti}, {Thuillot}, {van Reeven}, {Viala},
  {Abbas}, {Abreu Aramburu}, {Accart}, {Aguado}, {Allan}, {Allasia},
  {Altavilla}, {{\'A}lvarez}, {Alves}, {Anderson}, {Andrei}, {Anglada Varela},
  {Antiche}, {Antoja}, {Ant{\'o}n}, {Arcay}, {Atzei}, {Ayache}, {Bach},
  {Baker}, {Balaguer-N{\'u}{\~n}ez}, {Barache}, {Barata}, {Barbier}, {Barblan},
  {Baroni}, {Barrado y Navascu{\'e}s}, {Barros}, {Barstow}, {Becciani},
  {Bellazzini}, {Bellei}, {Bello Garc{\'\i}a}, {Belokurov}, {Bendjoya},
  {Berihuete}, {Bianchi}, {Bienaym{\'e}}, {Billebaud}, {Blagorodnova},
  {Blanco-Cuaresma}, {Boch}, {Bombrun}, {Borrachero}, {Bouquillon}, {Bourda},
  {Bouy}, {Bragaglia}, {Breddels}, {Brouillet}, {Br{\"u}semeister},
  {Bucciarelli}, {Budnik}, {Burgess}, {Burgon}, {Burlacu}, {Busonero}, {Buzzi},
  {Caffau}, {Cambras}, {Campbell}, {Cancelliere}, {Cantat-Gaudin}, {Carlucci},
  {Carrasco}, {Castellani}, {Charlot}, {Charnas}, {Charvet}, {Chassat},
  {Chiavassa}, {Clotet}, {Cocozza}, {Collins}, {Collins}, {Costigan}, {Crifo},
  {Cross}, {Crosta}, {Crowley}, {Dafonte}, {Damerdji}, {Dapergolas}, {David},
  {David}, {De Cat}, {de Felice}, {de Laverny}, {De Luise}, {De March}, {de
  Martino}, {de Souza}, {Debosscher}, {del Pozo}, {Delbo}, {Delgado},
  {Delgado}, {di Marco}, {Di Matteo}, {Diakite}, {Distefano}, {Dolding}, {Dos
  Anjos}, {Drazinos}, {Dur{\'a}n}, {Dzigan}, {Ecale}, {Edvardsson}, {Enke},
  {Erdmann}, {Escolar}, {Espina}, {Evans}, {Eynard Bontemps}, {Fabre},
  {Fabrizio}, {Faigler}, {Falc{\~a}o}, {Farr{\`a}s Casas}, {Faye}, {Federici},
  {Fedorets}, {Fern{\'a}ndez-Hern{\'a}ndez}, {Fernique}, {Fienga}, {Figueras},
  {Filippi}, {Findeisen}, {Fonti}, {Fouesneau}, {Fraile}, {Fraser}, {Fuchs},
  {Furnell}, {Gai}, {Galleti}, {Galluccio}, {Garabato}, {Garc{\'\i}a-Sedano},
  {Gar{\'e}}, {Garofalo}, {Garralda}, {Gavras}, {Gerssen}, {Geyer}, {Gilmore},
  {Girona}, {Giuffrida}, {Gomes}, {Gonz{\'a}lez-Marcos},
  {Gonz{\'a}lez-N{\'u}{\~n}ez}, {Gonz{\'a}lez-Vidal}, {Granvik}, {Guerrier},
  {Guillout}, {Guiraud}, {G{\'u}rpide}, {Guti{\'e}rrez-S{\'a}nchez}, {Guy},
  {Haigron}, {Hatzidimitriou}, {Haywood}, {Heiter}, {Helmi}, {Hobbs},
  {Hofmann}, {Holl}, {Holland}, {Hunt}, {Hypki}, {Icardi}, {Irwin}, {Jevardat
  de Fombelle}, {Jofr{\'e}}, {Jonker}, {Jorissen}, {Julbe}, {Karampelas},
  {Kochoska}, {Kohley}, {Kolenberg}, {Kontizas}, {Koposov}, {Kordopatis},
  {Koubsky}, {Kowalczyk}, {Krone-Martins}, {Kudryashova}, {Kull}, {Bachchan},
  {Lacoste-Seris}, {Lanza}, {Lavigne}, {Le Poncin-Lafitte}, {Lebreton},
  {Lebzelter}, {Leccia}, {Leclerc}, {Lecoeur-Taibi}, {Lemaitre}, {Lenhardt},
  {Leroux}, {Liao}, {Licata}, {Lindstr{\o}m}, {Lister}, {Livanou}, {Lobel},
  {L{\"o}ffler}, {L{\'o}pez}, {Lopez-Lozano}, {Lorenz}, {Loureiro},
  {MacDonald}, {Magalh{\~a}es Fernandes}, {Managau}, {Mann}, {Mantelet},
  {Marchal}, {Marchant}, {Marconi}, {Marie}, {Marinoni}, {Marrese},
  {Marschalk{\'o}}, {Marshall}, {Mart{\'\i}n-Fleitas}, {Martino}, {Mary},
  {Matijevi{\v{c}}}, {Mazeh}, {McMillan}, {Messina}, {Mestre}, {Michalik},
  {Millar}, {Miranda}, {Molina}, {Molinaro}, {Molinaro}, {Moln{\'a}r},
  {Moniez}, {Montegriffo}, {Monteiro}, {Mor}, {Mora}, {Morbidelli}, {Morel},
  {Morgenthaler}, {Morley}, {Morris}, {Mulone}, {Muraveva}, {Musella},
  {Narbonne}, {Nelemans}, {Nicastro}, {Noval}, {Ord{\'e}novic},
  {Ordieres-Mer{\'e}}, {Osborne}, {Pagani}, {Pagano}, {Pailler}, {Palacin},
  {Palaversa}, {Parsons}, {Paulsen}, {Pecoraro}, {Pedrosa}, {Pentik{\"a}inen},
  {Pereira}, {Pichon}, {Piersimoni}, {Pineau}, {Plachy}, {Plum}, {Poujoulet},
  {Pr{\v{s}}a}, {Pulone}, {Ragaini}, {Rago}, {Rambaux}, {Ramos-Lerate},
  {Ranalli}, {Rauw}, {Read}, {Regibo}, {Renk}, {Reyl{\'e}}, {Ribeiro},
  {Rimoldini}, {Ripepi}, {Riva}, {Rixon}, {Roelens}, {Romero-G{\'o}mez},
  {Rowell}, {Royer}, {Rudolph}, {Ruiz-Dern}, {Sadowski}, {Sagrist{\`a}
  Sell{\'e}s}, {Sahlmann}, {Salgado}, {Salguero}, {Sarasso}, {Savietto},
  {Schnorhk}, {Schultheis}, {Sciacca}, {Segol}, {Segovia}, {Segransan},
  {Serpell}, {Shih}, {Smareglia}, {Smart}, {Smith}, {Solano}, {Solitro},
  {Sordo}, {Soria Nieto}, {Souchay}, {Spagna}, {Spoto}, {Stampa}, {Steele},
  {Steidelm{\"u}ller}, {Stephenson}, {Stoev}, {Suess}, {S{\"u}veges}, {Surdej},
  {Szabados}, {Szegedi-Elek}, {Tapiador}, {Taris}, {Tauran}, {Taylor},
  {Teixeira}, {Terrett}, {Tingley}, {Trager}, {Turon}, {Ulla}, {Utrilla},
  {Valentini}, {van Elteren}, {Van Hemelryck}, {van Leeuwen}, {Varadi},
  {Vecchiato}, {Veljanoski}, {Via}, {Vicente}, {Vogt}, {Voss}, {Votruba},
  {Voutsinas}, {Walmsley}, {Weiler}, {Weingrill}, {Werner}, {Wevers},
  {Whitehead}, {Wyrzykowski}, {Yoldas}, {{\v{Z}}erjal}, {Zucker}, {Zurbach},
  {Zwitter}, {Alecu}, {Allen}, {Allende Prieto}, {Amorim},
  {Anglada-Escud{\'e}}, {Arsenijevic}, {Azaz}, {Balm}, {Beck}, {Bernstein},
  {Bigot}, {Bijaoui}, {Blasco}, {Bonfigli}, {Bono}, {Boudreault}, {Bressan},
  {Brown}, {Brunet}, {Bunclark}, {Buonanno}, {Butkevich}, {Carret}, {Carrion},
  {Chemin}, {Ch{\'e}reau}, {Corcione}, {Darmigny}, {de Boer}, {de Teodoro}, {de
  Zeeuw}, {Delle Luche}, {Domingues}, {Dubath}, {Fodor}, {Fr{\'e}zouls},
  {Fries}, {Fustes}, {Fyfe}, {Gallardo}, {Gallegos}, {Gardiol}, {Gebran},
  {Gomboc}, {G{\'o}mez}, {Grux}, {Gueguen}, {Heyrovsky}, {Hoar}, {Iannicola},
  {Isasi Parache}, {Janotto}, {Joliet}, {Jonckheere}, {Keil}, {Kim},
  {Klagyivik}, {Klar}, {Knude}, {Kochukhov}, {Kolka}, {Kos}, {Kutka}, {Lainey},
  {LeBouquin}, {Liu}, {Loreggia}, {Makarov}, {Marseille}, {Martayan},
  {Martinez-Rubi}, {Massart}, {Meynadier}, {Mignot}, {Munari}, {Nguyen},
  {Nordlander}, {Ocvirk}, {O'Flaherty}, {Olias Sanz}, {Ortiz}, {Osorio},
  {Oszkiewicz}, {Ouzounis}, {Palmer}, {Park}, {Pasquato}, {Peltzer}, {Peralta},
  {P{\'e}turaud}, {Pieniluoma}, {Pigozzi}, {Poels}, {Prat}, {Prod'homme},
  {Raison}, {Rebordao}, {Risquez}, {Rocca-Volmerange}, {Rosen}, {Ruiz-Fuertes},
  {Russo}, {Sembay}, {Serraller Vizcaino}, {Short}, {Siebert}, {Silva},
  {Sinachopoulos}, {Slezak}, {Soffel}, {Sosnowska}, {Strai{\v{z}}ys}, {ter
  Linden}, {Terrell}, {Theil}, {Tiede}, {Troisi}, {Tsalmantza}, {Tur},
  {Vaccari}, {Vachier}, {Valles}, {Van Hamme}, {Veltz}, {Virtanen}, {Wallut},
  {Wichmann}, {Wilkinson}, {Ziaeepour}, \&
  {Zschocke}}]{Gaia-Collaboration:2016uw}
{Gaia Collaboration}, {Prusti}, T., {de Bruijne}, J.~H.~J., {et~al.} 2016,
  \aap, 595, A1

\bibitem[{{Gallego-Cano} {et~al.}(2020){Gallego-Cano}, {Sch{\"o}del},
  {Nogueras-Lara}, {Dong}, {Shahzamanian}, {Fritz}, {Gallego-Calvente}, \&
  {Neumayer}}]{gallego-cano2019}
{Gallego-Cano}, E., {Sch{\"o}del}, R., {Nogueras-Lara}, F., {et~al.} 2020,
  \aap, 634, A71

\bibitem[{{Girardi}(2016)}]{Girardi:2016fk}
{Girardi}, L. 2016, \araa, 54, 95

\bibitem[{{Gravity Collaboration} {et~al.}(2018){Gravity Collaboration},
  {Abuter}, {Amorim}, {Anugu}, {Baub{\"o}ck}, {Benisty}, {Berger}, {Blind},
  {Bonnet}, {Brandner}, {Buron}, {Collin}, {Chapron}, {Cl{\'e}net}, {Coud{\'e}
  Du Foresto}, {de Zeeuw}, {Deen}, {Delplancke-Str{\"o}bele}, {Dembet},
  {Dexter}, {Duvert}, {Eckart}, {Eisenhauer}, {Finger}, {F{\"o}rster
  Schreiber}, {F{\'e}dou}, {Garcia}, {Garcia Lopez}, {Gao}, {Gendron},
  {Genzel}, {Gillessen}, {Gordo}, {Habibi}, {Haubois}, {Haug}, {Hau{\ss}mann},
  {Henning}, {Hippler}, {Horrobin}, {Hubert}, {Hubin}, {Jimenez Rosales},
  {Jochum}, {Jocou}, {Kaufer}, {Kellner}, {Kendrew}, {Kervella}, {Kok},
  {Kulas}, {Lacour}, {Lapeyr{\`e}re}, {Lazareff}, {Le Bouquin}, {L{\'e}na},
  {Lippa}, {Lenzen}, {M{\'e}rand}, {M{\"u}ler}, {Neumann}, {Ott}, {Palanca},
  {Paumard}, {Pasquini}, {Perraut}, {Perrin}, {Pfuhl}, {Plewa}, {Rabien},
  {Ram{\'{\i}}rez}, {Ramos}, {Rau}, {Rodr{\'{\i}}guez-Coira}, {Rohloff},
  {Rousset}, {Sanchez-Bermudez}, {Scheithauer}, {Sch{\"o}ller}, {Schuler},
  {Spyromilio}, {Straub}, {Straubmeier}, {Sturm}, {Tacconi}, {Tristram},
  {Vincent}, {von Fellenberg}, {Wank}, {Waisberg}, {Widmann}, {Wieprecht},
  {Wiest}, {Wiezorrek}, {Woillez}, {Yazici}, {Ziegler}, \&
  {Zins}}]{Gravity-Collaboration:2018aa}
{Gravity Collaboration}, {Abuter}, R., {Amorim}, A., {et~al.} 2018, \aap, 615,
  L15

\bibitem[{{Groenewegen} {et~al.}(2008){Groenewegen}, {Udalski}, \&
  {Bono}}]{Groenewegen:2008by}
{Groenewegen}, M.~A.~T., {Udalski}, A., \& {Bono}, G. 2008, \aap, 481, 441

\bibitem[{{Henshaw} {et~al.}(2022){Henshaw}, {Barnes}, {Battersby}, {Ginsburg},
  {Sormani}, \& {Walker}}]{Henshaw:2022vl}
{Henshaw}, J.~D., {Barnes}, A.~T., {Battersby}, C., {et~al.} 2022, arXiv
  e-prints, arXiv:2203.11223

\bibitem[{{Launhardt} {et~al.}(2002){Launhardt}, {Zylka}, \&
  {Mezger}}]{Launhardt:2002nx}
{Launhardt}, R., {Zylka}, R., \& {Mezger}, P.~G. 2002, \aap, 384, 112

\bibitem[{{Libralato} {et~al.}(2021){Libralato}, {Lennon}, {Bellini}, {van der
  Marel}, {Clark}, {Najarro}, {Patrick}, {Anderson}, {Bedin}, {Crowther}, {de
  Mink}, {Evans}, {Platais}, {Sabbi}, \& {Sohn}}]{Libralato:2021td}
{Libralato}, M., {Lennon}, D.~J., {Bellini}, A., {et~al.} 2021, \mnras, 500,
  3213

\bibitem[{{Mart{\'\i}nez-Arranz} {et~al.}(2022){Mart{\'\i}nez-Arranz},
  {Sch{\"o}del}, {Nogueras-Lara}, \& {Shahzamanian}}]{Martinez-Arranz:2022uf}
{Mart{\'\i}nez-Arranz}, {\'A}., {Sch{\"o}del}, R., {Nogueras-Lara}, F., \&
  {Shahzamanian}, B. 2022, \aap, 660, L3

\bibitem[{{Matsunaga} {et~al.}(2011){Matsunaga}, {Kawadu}, {Nishiyama},
  {Nagayama}, {Kobayashi}, {Tamura}, {Bono}, {Feast}, \&
  {Nagata}}]{Matsunaga:2011uq}
{Matsunaga}, N., {Kawadu}, T., {Nishiyama}, S., {et~al.} 2011, \nat, 477, 188

\bibitem[{{Mezger} {et~al.}(1996){Mezger}, {Duschl}, \&
  {Zylka}}]{Mezger:1996uq}
{Mezger}, P.~G., {Duschl}, W.~J., \& {Zylka}, R. 1996, \aapr, 7, 289

\bibitem[{{Nishiyama} {et~al.}(2006){Nishiyama}, {Nagata}, {Sato}, {Kato},
  {Nagayama}, {Kusakabe}, {Matsunaga}, {Naoi}, {Sugitani}, \&
  {Tamura}}]{Nishiyama:2006ai}
{Nishiyama}, S., {Nagata}, T., {Sato}, S., {et~al.} 2006, \apj, 647, 1093

\bibitem[{{Nishiyama} {et~al.}(2008){Nishiyama}, {Nagata}, {Tamura}, {Kandori},
  {Hatano}, {Sato}, \& {Sugitani}}]{Nishiyama:2008qa}
{Nishiyama}, S., {Nagata}, T., {Tamura}, M., {et~al.} 2008, \apj, 680, 1174

\bibitem[{{Nogueras-Lara}(2022)}]{Nogueras-Lara:2022tp}
{Nogueras-Lara}, F. 2022, \aap, 666, A72

\bibitem[{{Nogueras-Lara} {et~al.}(2018){Nogueras-Lara}, {Gallego-Calvente},
  {Dong}, {Gallego-Cano}, {Girard}, {Hilker}, {de Zeeuw}, {Feldmeier-Krause},
  {Nishiyama}, {Najarro}, {Neumayer}, \& {Sch{\"o}del}}]{Nogueras-Lara:2018aa}
{Nogueras-Lara}, F., {Gallego-Calvente}, A.~T., {Dong}, H., {et~al.} 2018,
  \aap, 610, A83

\bibitem[{{Nogueras-Lara} {et~al.}(2019){Nogueras-Lara}, {Sch{\"o}del},
  {Gallego-Calvente}, {Dong}, {Gallego-Cano}, {Shahzamanian}, {Girard},
  {Nishiyama}, {Najarro}, \& {Neumayer}}]{Nogueras-Lara:2019aa}
{Nogueras-Lara}, F., {Sch{\"o}del}, R., {Gallego-Calvente}, A.~T., {et~al.}
  2019, \aap, 631, A20

\bibitem[{{Nogueras-Lara} {et~al.}(2020{\natexlab{a}}){Nogueras-Lara},
  {Sch{\"o}del}, {Gallego-Calvente}, {Gallego-Cano}, {Shahzamanian}, {Dong},
  {Neumayer}, {Hilker}, {Najarro}, {Nishiyama}, {Feldmeier-Krause}, {Girard},
  \& {Cassisi}}]{Nogueras-Lara:2019ad}
{Nogueras-Lara}, F., {Sch{\"o}del}, R., {Gallego-Calvente}, A.~T., {et~al.}
  2020{\natexlab{a}}, Nature Astronomy, 4, 377

\bibitem[{{Nogueras-Lara} {et~al.}(2021{\natexlab{a}}){Nogueras-Lara},
  {Sch{\"o}del}, \& {Neumayer}}]{Nogueras-Lara:2021uz}
{Nogueras-Lara}, F., {Sch{\"o}del}, R., \& {Neumayer}, N. 2021{\natexlab{a}},
  \aap, 653, A33

\bibitem[{{Nogueras-Lara} {et~al.}(2021{\natexlab{b}}){Nogueras-Lara},
  {Sch{\"o}del}, \& {Neumayer}}]{Nogueras-Lara:2021wj}
{Nogueras-Lara}, F., {Sch{\"o}del}, R., \& {Neumayer}, N. 2021{\natexlab{b}},
  \aap, 653, A133

\bibitem[{{Nogueras-Lara} {et~al.}(2021{\natexlab{c}}){Nogueras-Lara},
  {Sch{\"o}del}, \& {Neumayer}}]{Nogueras-Lara:2021wm}
{Nogueras-Lara}, F., {Sch{\"o}del}, R., \& {Neumayer}, N. 2021{\natexlab{c}},
  \apj, 920, 97

\bibitem[{{Nogueras-Lara} {et~al.}(2022){Nogueras-Lara}, {Sch{\"o}del}, \&
  {Neumayer}}]{Nogueras-Lara:2022ua}
{Nogueras-Lara}, F., {Sch{\"o}del}, R., \& {Neumayer}, N. 2022, Nature
  Astronomy, 6, 1178

\bibitem[{{Nogueras-Lara} {et~al.}(2020{\natexlab{b}}){Nogueras-Lara},
  {Sch{\"o}del}, {Neumayer}, {Gallego-Cano}, {Shahzamanian},
  {Gallego-Calvente}, \& {Najarro}}]{Nogueras-Lara:2020aa}
{Nogueras-Lara}, F., {Sch{\"o}del}, R., {Neumayer}, N., {et~al.}
  2020{\natexlab{b}}, \aap, 641, A141

\bibitem[{{Nogueras-Lara} {et~al.}(2021{\natexlab{d}}){Nogueras-Lara},
  {Sch{\"o}del}, {Neumayer}, \& {Schultheis}}]{Nogueras-Lara:2021uq}
{Nogueras-Lara}, F., {Sch{\"o}del}, R., {Neumayer}, N., \& {Schultheis}, M.
  2021{\natexlab{d}}, \aap, 647, L6

\bibitem[{Pedregosa {et~al.}(2011)Pedregosa, Varoquaux, Gramfort, Michel,
  Thirion, Grisel, Blondel, Prettenhofer, Weiss, Dubourg, Vanderplas, Passos,
  Cournapeau, Brucher, Perrot, \& Duchesnay}]{Pedregosa:2011aa}
Pedregosa, F., Varoquaux, G., Gramfort, A., {et~al.} 2011, Journal of Machine
  Learning Research, 12, 2825

\bibitem[{{Sch{\"o}del} {et~al.}(2014){Sch{\"o}del}, {Feldmeier}, {Kunneriath},
  {Stolovy}, {Neumayer}, {Amaro-Seoane}, \& {Nishiyama}}]{Schodel:2014fk}
{Sch{\"o}del}, R., {Feldmeier}, A., {Kunneriath}, D., {et~al.} 2014, \aap, 566,
  A47

\bibitem[{{Sch{\"o}nrich} {et~al.}(2015){Sch{\"o}nrich}, {Aumer}, \&
  {Sale}}]{2015ApJ...812L..21S}
{Sch{\"o}nrich}, R., {Aumer}, M., \& {Sale}, S.~E. 2015, \apjl, 812, L21

\bibitem[{{Schultheis} {et~al.}(2021){Schultheis}, {Fritz}, {Nandakumar},
  {Rojas-Arriagada}, {Nogueras-Lara}, {Feldmeier-Krause}, {Gerhard},
  {Neumayer}, {Patrick}, {Prieto}, {Sch{\"o}del}, {Mastrobuono-Battisti}, \&
  {Sormani}}]{Schultheis:2021wf}
{Schultheis}, M., {Fritz}, T.~K., {Nandakumar}, G., {et~al.} 2021, \aap, 650,
  A191

\bibitem[{Schwarz(1978)}]{Schwarz:1978aa}
Schwarz, G. 1978, The Annals of Statistics, 6, 461

\bibitem[{{Shahzamanian} {et~al.}(2022){Shahzamanian}, {Sch{\"o}del},
  {Nogueras-Lara}, {Mart{\'\i}nez-Arranz}, {Sormani}, {Gallego-Calvente},
  {Gallego-Cano}, \& {Alburai}}]{Shahzamanian:2021wu}
{Shahzamanian}, B., {Sch{\"o}del}, R., {Nogueras-Lara}, F., {et~al.} 2022,
  \aap, 662, A11

\bibitem[{{Sormani} {et~al.}(2020){Sormani}, {Magorrian}, {Nogueras-Lara},
  {Neumayer}, {Sch{\"o}nrich}, {Klessen}, \&
  {Mastrobuono-Battisti}}]{Sormani:2020aa}
{Sormani}, M.~C., {Magorrian}, J., {Nogueras-Lara}, F., {et~al.} 2020, \mnras,
  499, 7

\bibitem[{{Sormani} {et~al.}(2022){Sormani}, {Sanders}, {Fritz}, {Smith},
  {Gerhard}, {Sch{\"o}del}, {Magorrian}, {Neumayer}, {Nogueras-Lara},
  {Feldmeier-Krause}, {Mastrobuono-Battisti}, {Schultheis}, {Shahzamanian},
  {Vasiliev}, {Klessen}, {Lucas}, \& {Minniti}}]{Sormani:2022wv}
{Sormani}, M.~C., {Sanders}, J.~L., {Fritz}, T.~K., {et~al.} 2022, \mnras, 512,
  1857

\bibitem[{{Sormani} {et~al.}(2018){Sormani}, {Tre{\ss}}, {Ridley}, {Glover},
  {Klessen}, {Binney}, {Magorrian}, \& {Smith}}]{Sormani:2018aa}
{Sormani}, M.~C., {Tre{\ss}}, R.~G., {Ridley}, M., {et~al.} 2018, \mnras, 475,
  2383

\end{thebibliography}
\end{document}